\documentclass[10pt,conference]{IEEEtran}
\IEEEoverridecommandlockouts
\usepackage[colorlinks=true,linkcolor=blue,urlcolor=blue,citecolor=blue,anchorcolor=blue]{hyperref}
\usepackage[numbers,sort&compress,square]{natbib}
\usepackage{amsmath,amssymb,amsfonts}
\usepackage{algorithmic}
\usepackage{graphicx}
\usepackage{booktabs}
\usepackage{textcomp}
\usepackage{xcolor}
\def\BibTeX{{\rm B\kern-.05em{\sc i\kern-.025em b}\kern-.08em
    T\kern-.1667em\lower.7ex\hbox{E}\kern-.125emX}}

\begin{document}

\title{Which contributions count?\\ 
Analysis of attribution in open source}

\author
{\IEEEauthorblockN{Jean-Gabriel Young}
\IEEEauthorblockA{
jean-gabriel.young@uvm.edu\\
Department of Computer Science\\
Vermont Complex Systems Center\\
University of Vermont\\
Burlington VT, USA }\\
\IEEEauthorblockN{Milo Z. Trujillo}
\IEEEauthorblockA{
milo.trujillo@uvm.edu\\
Vermont Complex Systems Center\\
University of Vermont\\
Burlington VT, USA }\\
\and
\IEEEauthorblockN{Amanda Casari}
\IEEEauthorblockA{
amcasari@google.com\\
Open Source Programs Office, Google\\
Kirkland, WA, USA}\\
\IEEEauthorblockN{Laurent H\'ebert-Dufresne}
\IEEEauthorblockA{
laurent.hebert-dufresne@uvm.edu\\
Department of Computer Science\\
Vermont Complex Systems Center\\
University of Vermont\\
Burlington VT, USA }
\and
\IEEEauthorblockN{Katie McLaughlin}
\IEEEauthorblockA{
glasnt@google.com\\
Open Source Programs Office, Google\\
Sydney, New South Wales, Australia}\\
\IEEEauthorblockN{James P. Bagrow}
\IEEEauthorblockA{
james.bagrow@uvm.edu\\
Department of Mathematics \& Statistics\\
Vermont Complex Systems Center\\
University of Vermont\\
Burlington VT, USA }}

\maketitle

\begin{abstract}
Open source software projects usually acknowledge contributions with text files, websites, and other idiosyncratic methods.
These data sources are hard to mine, which is why contributorship is most frequently measured through changes to repositories, such as commits, pushes, or patches.
Recently, some open source projects have taken to recording contributor actions with standardized systems; this opens up a unique opportunity to understand how community-generated notions of contributorship map onto codebases as the measure of contribution.
Here, we characterize contributor acknowledgment models in open source by analyzing thousands of projects that use a model called All Contributors to acknowledge diverse contributions like outreach, finance, infrastructure, and community management. 
We analyze the life cycle of projects through this model's lens and contrast its representation of contributorship with the picture given by other methods of acknowledgment, including GitHub's top committers indicator and contributions derived from actions taken on the platform.
We find that community-generated systems of contribution acknowledgment make work like idea generation or bug finding more visible, which generates a more extensive picture of collaboration.
Further, we find that models requiring explicit attribution lead to more clearly defined boundaries around what is and what is not a contribution.
\end{abstract}

\begin{IEEEkeywords}
open source software, contributions, teams, github
\end{IEEEkeywords}

\section{Introduction}

Producing software is a multifaceted activity that requires expertise beyond just writing code.
Besides developers, tech organizations employ workers in diverse roles such as testing, product management, human resources, sales, and so forth \cite{ramin2020more}.
All of these roles contribute to the core product \cite{brooks1995mythical}, if indirectly, and are recognized as such.

Unlike industrial software, open source software (OSS) is often developed under a non-traditional structure and, as a result, is seen as the product of teams composed almost exclusively of developers~\cite{eghbal2016roads}.
This picture is of course incomplete at best, as is well-known by those involved with OSS \cite{eghbal2020working}.
While young projects can thrive under the guidance of lone developers or small unsupported teams, more mature projects usually benefit from contributions to the project that transcend code \cite{eghbal2020working}.
These non-code contributions may include, for example: moderating communication channels associated with the project or its issue tracker(s), fielding questions, outreach, infrastructure, governance, funding acquisition, documentation, or even mere \emph{attention}~\cite{middleton2018contributions,eghbal2020working}---these contributions are all crucial determinants of a project's continued success \cite{alliez2019attributing,fang2020need,eghbal2020working}.

The predominant incentive structure under which OSS operates makes all non-code contributions practically invisible to outsiders.
Someone who is well acquainted with a community might be highly aware of who spent a lot of time moderating a community, developing the project's road map, or managing the queue of issues.
However, this information is rarely recorded in a standardized way, particularly within software repositories.
Instead, OSS projects tend to acknowledge contributions in ad-hoc ways, with mechanisms that include, for example:  flat credit files, acknowledgments appearing on a project's website,  challenge coins handed out to contributing community members \cite{beeware}, or even academic papers providing citable artifacts and a snapshot of a project at a particular moment in time.
These models of acknowledgments are far too diverse and variable to mine at scale.
Further, they are not always attached to the software project itself.
As a result, when mining software repositories, we are more or less reduced to defining contributorship in terms of changes to the repositories, whether it be at the levels of lines of codes, commits, pull or merge requests, or patches.

Definitions of contributorship that emphasize code changes are far from perfect.
First, strictly focusing on the data mining perspective, code as the measure of contribution tends to provide an inaccurate picture of OSS projects.
For example, two definitions can be seemingly in agreement---say, having committed code or opened a pull request---yet lead to different conclusions about how OSS is made \cite{bertoncello2020pull}.
Commit data may be erased or altered during code merges and revisions~\cite{pinto2018leaving}, making decentralized version control logs unreliable for measuring contributorships~\cite{bird2009promises}.
As we have already mentioned, contributorship, defined as the changing of code, does not cover the complete spectrum of contributions \cite{gousios2008measuring,alliez2019attributing}.
Second, contributorship-as-code does not always map to a meaningful notion of contribution.
Not only do developers routinely contribute code without feeling a sense of ownership or deeming themselves contributor to a project  \cite{pinto2016more,eghbal2020working}, but increasingly large parts of code changes are implemented by bots that act as users~\cite{wessel2018power}.
Finally, there are pronounced gender \cite{terrell2017gender,robles2016women,kuechler2012gender} and geographical \cite{furtado2020successful} differences in the way people contribute to OSS---such that ``commit as contribution'' is not likely to capture a representative cross-section of contributors.

Now, given that: (1) ``commit as contribution'' is a flawed model, and (2) more representative acknowledgment models are hard to mine due to a lack of standardization across projects, one might wonder: \emph{Are we destined to mischaracterize contributorship to OSS projects in all our large-scale analyses?}

Fortunately, new tools \cite{mclaughlin2016acknowledging}, and models \cite{allcontributors,ramin2020more} could help us derive a more accurate picture of who makes OSS and how.
These tools and models aim to provide more systematic and complete coverage of the types of contributions routinely made to OSS.
For example, tools like \texttt{octohatrack}~\cite{octohatrack} and \texttt{name-your-contributors} \cite{nameyourcontributors} can generate lists of all the users that have interacted with a GitHub project, including those who opened issues, created pull requests, or discussed issues.
Models like the All Contributors (AC) project aim to standardize credit files~\cite{allcontributors}, and recommend that all the contributors be credited together with the types of contribution they have made, such as coding, code review, translation work, or financial support.
Other models, like \texttt{gitmoji} \cite{gitmoji}, classify commits by the type of contribution they make to a project.
And many groups of researchers have proposed to record the role of team members alongside codebases \cite{ramin2020more,casari2021open}, or as metadata \cite{alliez2019attributing}.

These tools and models have yet to permeate OSS, but some of them have gained a small and dedicated following.
As a result, several projects already have detailed and parsable contribution histories going back many years.
To the best of our knowledge, this is the first time where we have access to a sizable, easily mineable, and detailed corpus of contributorship data in OSS, where contributorship is established by the project members themselves instead of through external analysis conventions.

In this paper, we characterize the use of these attribution systems on GitHub.
We organize our analysis around a series of research questions. First, we ask:\
\begin{itemize}
    \item {RQ1}: \emph{How do models of contributorship acknowledgment differ?}
    \item {RQ2}: \emph{How much information is missed by focusing on repository changes as the model of contribution?}
\end{itemize}
Having understood the kind of information given by various systems of contributions acknowledgment, we then focus on:\
\begin{itemize}
    \item {RQ3}: \emph{How do contribution to open source projects evolve as they age?}
    \item {RQ4}: \emph{Can we classify projects based on patterns of contributions?}
\end{itemize}

We discuss our answers to these questions in detail in Sec.~\ref{sec:results} and \ref{sec:discussion}, but we highlight three key findings here:
\begin{enumerate}
    \item \textbf{Generosity}. Models of contribution acknowledgment that require manual attribution of credits recognize fewer contributors than other models but that they can cover work not captured by other methods (even by proxy).
    \item \textbf{Large blind spots}. Across OSS, the majority of contributors are acknowledged by simple methods like code commits, but these methods may still omit the majority of contributors to \emph{individual} projects.
    \item \textbf{Typical life-cycle}. As OSS projects grow in age and popularity, contributors tend to spread more uniformly across types of contributions. We do not see a similar trend towards uniformity for tasks, in the sense that some projects see an increased concentration towards a few types of tasks (like code), while others grow in all areas simultaneously. 
\end{enumerate}

Together, the answers to these research questions show that we need to rethink how we define contributorship when mining software repositories.

\section{Background}
\label{sec:background}

We distinguish four models of contributorship acknowledgment.

First, we define \textbf{platform-attributed contributors} as the contributors that a user browsing GitHub can easily and reliably see.
By default, these contributors are the top 100 users who have created the most commits in a repository, which may (and often does) include bots.
The list can be re-ordered to show deletion or addition of code, but we will stick to commits as it is a more agreed-upon notion---if imperfect~\cite{bertoncello2020pull}---of contribution.
This type of contribution emphasizes code above all.

Second, we define \textbf{contributors identified by automated tools} as those contributors that are not \emph{necessarily} credited for work on a project, but that can nonetheless be identified programmatically \emph{for any project}.
In other words, data about these contributors is already available as a by-product of the development process, but it might not be highlighted by default.
A good example would be the contributors extracted by \texttt{octohatrack}~\cite{octohatrack}, which uses the GitHub Event Stream (GES) to construct a list of all the users having interacted with a repository.
Other methods of this type would include joining mailing lists or answers pulled from Q\&A sites with repository information.

Third, we define \textbf{contributors identified with taxonomies} as contributors who are credited through formatted files following some prescribed ``standard.''
These taxonomies are relatively recent additions to OSS, so there are few proposed models, fewer implementations, and no ratified standards.
In fact, we are aware of only one project that has gained traction thus far, the All Contributor (AC) model, whose traces can be found in roughly $15\,000$ repositories at the time of writing.
Repositories following this model contain a JSON file in the root folder (\texttt{.all-contributorsrc}).
The file lists all the users who have contributed to the project together with the types of contributions they have made.
These contributions are selected from a standardized list of 32 types of contributions emphasizing \emph{artifacts} (e.g., ``created documentation'')  and \emph{activities} (e.g., ``does business development'') and are attributed manually, oftentimes with the help of a bot to edit the file.
We note that other projects have attempted similar taxonomies; we focus on AC because these projects either do not have a broad notion of contributorship (e.g., \texttt{gitmoji}  only classifies code commits but does not credit non-code work) or have not yet gained significant traction (e.g., The Software Credit Ontology \cite{nieuwpoort2016software}).

Fourth and finally, we define \textbf{contributors identified by ad hoc methods} as those contributors that can be identified by parsing non-standardized data sources.
This includes contributors appearing on websites (e.g., a board of directors), in text files where the contribution made is not specified, in the documentation or the license files of projects, etc.
These sources of data require massive amounts of work to mine \cite{milewicz2019characterizing} but can provide valuable information nonetheless.

\section{Methods}

\subsection{Data pipeline: lists of contributors}
Due to the difficulty of mining ad hoc attribution data, we focused our analysis on the first three types of contributorship acknowledgment models and defined our sample as all the projects that implemented the AC model.

To mine this sample, we first generated an initial list of repositories by querying the GitHub API for repositories that had a \texttt{.all-contributorsrc} file in the root folder.
We obtained 6\,349 repositories in this way, down from 14\,191 repository in which the file appeared in included folders.
This initial filter selected for intentional use of the AC model.
We then downloaded the  \texttt{.all-contributorsrc} files of these projects and discarded those that had incorrect formatting (625 invalid JSON files) or recorded no contributions (234 files).
This left us with a starting list of 5\,490 repositories.

\begin{figure}
    \centering
    \includegraphics[width=\linewidth]{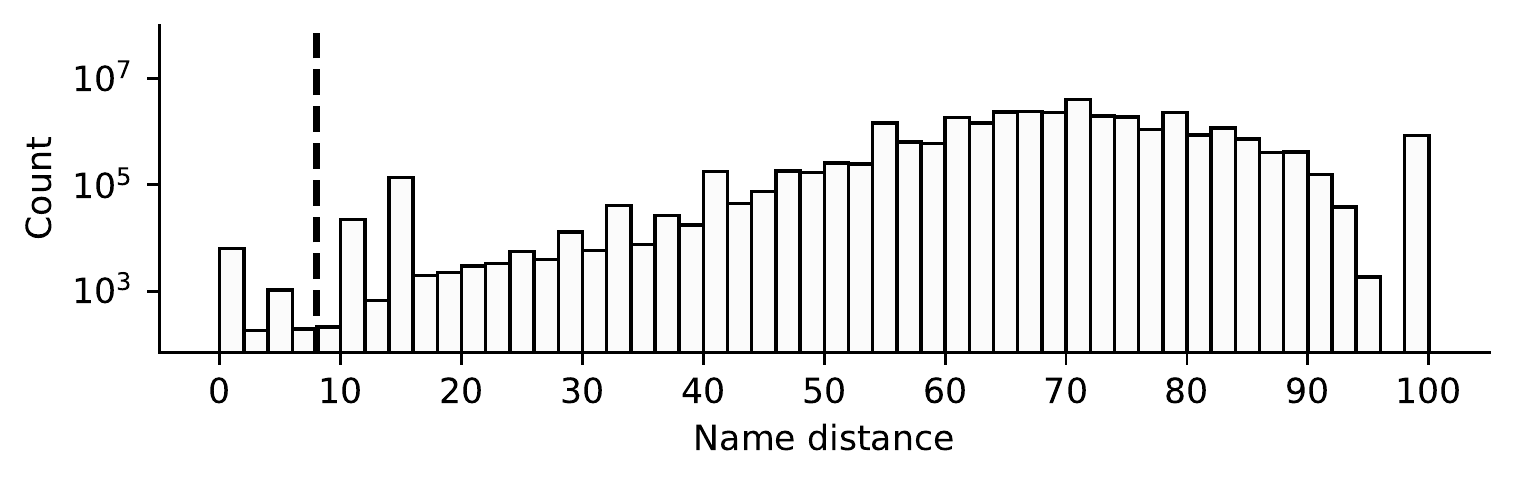}
    \caption{Detecting potential duplicates with repository names. We computed the distance between repositories' names with a modified Levenshtein Distance, where 0 indicates a perfect match and 100 a maximally distanced pair of names. We clustered the repositories with DBSCAN, treating pairs of repositories with names at a distance of 8 or less as potential duplicates. This yielded 28 clusters of more than one repository, accounting for 291 potential duplicates. The threshold was chosen by inspecting the clusters to look for a gap in the histogram of distances above.}
    \label{fig:name_clustering}
\end{figure}
Many of these repositories were inactive copies of others in our dataset \cite{kalliamvakou2016depth}, so we further filtered the list to keep only the repository that was the ``main development branch.''
To do this, we began by creating a list of potential duplicates by finding repositories in our dataset that:
\begin{enumerate}
    \item were the fork of another repository in our dataset, as identified by the GitHub API; 
    \item or were identified as a duplicate by Spinnelis and collaborators \cite{spinellis2020dataset};
    \item or fuzzily matched other repositories in our sample by name (using a modified Levenshtein distance to compute the similarity of names, followed by clustering the names with DBSCAN~\cite{pedregosa2011scikit}, see Fig.~\ref{fig:name_clustering});
    \item or had inconsistent project ownership information in the \texttt{.all-contributorsrc} file (project owners and names must appear in the file and might not match the actual repository from which we downloaded the file).
\end{enumerate}
The first criterion allowed us to identify potential copies created through the GitHub interface (18 pairs of repositories); the second gave us a (slightly outdated) list of potential copies by using indicators like the proximity of commit history (95 candidates); the third returned an up-to-date but incomplete coverage of possible duplicate (291 candidates), and the fourth gave a strong signal that the repository might have been copied from another without updating the files (2\,643 candidates).
Accounting for overlaps between these candidates, we were able to identify 2\,789 problematic repositories.
Since none of these four criteria could \emph{reliably} tell us whether a repository should remain in our dataset,\footnotemark\  we defaulted to removing these repositories from our sample but manually retained 142 (e.g., the ``true'' repository in each cluster of names or all of them in the match was spurious; repositories from the Spinnelis dataset \cite{spinellis2020dataset} that were valid, etc.).
We were left with a final sample of 2\,855 projects after removing spurious repositories.
\footnotetext{For example, a fork of a template might be the relevant repository instead of its parent; matching names do not necessarily imply matching content; a fork might lead to two distinct but active projects; repositories have aliases.}

To complement this sample, we gathered data for the two other models of contributorship acknowledgment discussed in Sec.~\ref{sec:background}.
For each project in our final sample, we retrieved a complete list of contributors with \texttt{octohatrack}, a Python package that returns a list of all the users that interacted with a repository together with the types of interaction they had with that repository, as recorded in the GES.
Further, we directly queried the GitHub API to retrieve a list of the top 100 contributors as defined by GitHub.
These two sources of data credited many bots, which we removed to focus on human contributors.
Naming conventions made this task relatively straightforward; we removed all users whose login ended in \texttt{[bot]} and \texttt{-bot}, together with a small curated list of bots whose names did not include these substrings (e.g., `renovate' or `all-contributors').
This filter allowed us to identify  7\,066 instances of bots contributing to a project in our sample.
After applying the filter, we ended up with a dataset comprising 142\,599 projects--contributors pairs.

As a final step, we gathered meta-data about these repositories, including the primary programming language of a repository (as determined by GitHub), their current popularity as measured by stars and forks, their velocity as quantified by the number of commits, and their community score, defined as the number of completed items on their community profile (which includes items like having a \texttt{README} file or a pull request template).

\subsection{Data pipeline: contribution history}

To gain further insights into the evolution of contributorship, we cloned the 100 projects with the largest number of contributors according to All Contributors and parsed their history.
This limited sample allowed us to focus on projects with a long history and large teams (the sample size of 100 projects is arbitrary and was selected before analyzing the data).
For each project, we then replayed the evolution of the  \texttt{.all-contributorsrc}  file starting when it was first created, using the version control information, and computed summary statistics of the contributions declared in each version of the file.
From these, we developed time series showing the evolution of these statistics.
We ignored versions of the file that did not respect the AC format description (e.g., written in invalid JSON or recording no contributor roles).
Note that we only used this restricted sample for RQ3 (results shown in Fig.~\ref{fig:growth}); all the other results were computed with the entire sample of 2\,855 projects.

\subsection{Coarse-grained contributions}

The types of contributions identified by AC and the GitHub API are somewhat granular.
For example, AC distinguishes ``plug-ins'' and ``code'' as separate types of contributions, while the GES (as mined by \texttt{octohatrack}) records events such as `PublicEvent' (making a private repository public) that do not occur often in the lifetime of projects.
Such granular data is not conducive to meaningful analysis of contributorship \emph{across} many repositories, due to repository-to-repository variations in the way contributions are acknowledged, and the use of the GitHub platform itself (say workflows centered around issues and pull requests versus commits,  periodic clones from a remote mirror, etc. \cite{kalliamvakou2014promises}).
To reduce the effect of such variations, we regrouped contributions into coarser categories, see Table~\ref{tab:coarse-graining-ac} and \ref{tab:coarse-graining-ohr}.
For AC, we obtained 5 coarse categories, down from 32, and for the GES we obtained 5 (down from 13).
When we applied this coarse-graining to the contribution data, we mapped non-standard categories to the best of our ability (e.g., we treated ``marketing'' as a ``business'' contribution). We ignored contributions that did not fit in the standard AC model in any way (e.g., ``discord'' or ``former-staff''). 
Unless specified, all our results are computed with the coarsened categories.

\renewcommand{\arraystretch}{1.2}
\begin{table}
\caption{Coarse-graining of the  All Contributors  taxonomy}
\label{tab:coarse-graining-ac}
\begin{tabular}{l|l}
    \toprule
    Coarse\! contribution\! & AC contribution\footnotemark\\
    \midrule
    Artifacts    &  \texttt{a11y}, \texttt{code}, \texttt{data}, \texttt{doc}, \texttt{design}, \texttt{plugin}\\
    &  \texttt{tool}, \texttt{translation}, \texttt{tests}, \texttt{userTesting}\\
    Education  & \texttt{audio},  \texttt{blog}, \texttt{content}, \texttt{example}\\
    \null\ \ \& Outreach& \texttt{eventOrganizing}, \texttt{mentoring}, \texttt{question}\\
            &  \texttt{talk}, \texttt{tutorial}, \texttt{video}\\
    Lead &  \texttt{business}, \texttt{financial}, \texttt{fundingFinding}\\
    &\texttt{ideas}, \texttt{projectManagement}, \texttt{research}\\
    Maintenance & \texttt{bugs}, \texttt{maintenance}, \texttt{review}\\
    Support & \texttt{infra}, \texttt{platform}, \texttt{security}\\
    \bottomrule
\end{tabular}
\end{table}
\footnotetext{\url{https://allcontributors.org/docs/en/emoji-key}}
\begin{table}
\vspace{\baselineskip}
\caption{Coarse-graining of GitHub Events Stream}
\label{tab:coarse-graining-ohr}
\begin{tabular}{l|l}
    \toprule
    Coarse contribution & Event\footnotemark \\
    \midrule
    Code & \texttt{PushEvent}, \texttt{PullRequestEvent} \\
    Code Review & \texttt{CommitCommentEvent}\\
                    & \texttt{PullRequestReviewEvent} \\
                & \texttt{PullRequestReviewCommentEvent} \\
    Issues &  \texttt{IssueCommentEvent}, \texttt{IssuesEvent} \\
    Maintenance& \texttt{CreateEvent}, \texttt{DeleteEvent}, \texttt{MemberEvent} \\
                & \texttt{PublicEvent}, \texttt{ReleaseEvent} \\
    Wiki/Docs& \texttt{GollumEvent} \\
    \bottomrule
\end{tabular}
\end{table}
\footnotetext{\url{https://docs.github.com/developers/webhooks-and-events/github-event-types}}

\subsection{Reproducibility}

Our analysis and code are available online\footnote{\url{https://doi.org/10.6084/m9.figshare.13966898.v1}}, and contains all the information needed to reproduce our analysis, to the extent made possible by best ethical practices.

\section{Results}
\label{sec:results}

We organize our results according to our four research questions.

\begin{figure}
\centering
\includegraphics[width=0.8\linewidth]{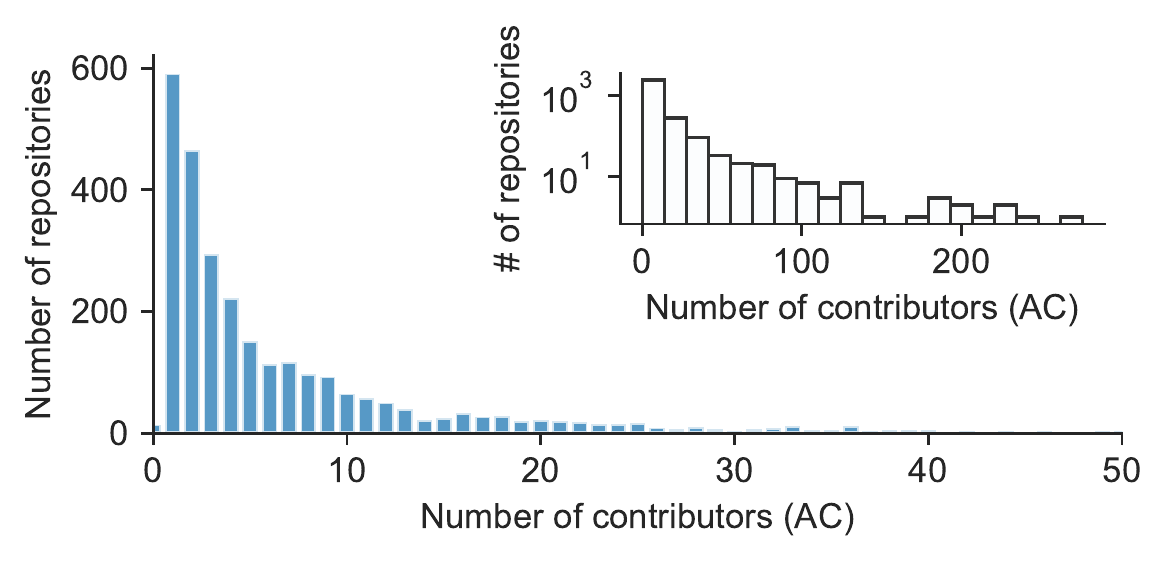}
\includegraphics[width=0.8\linewidth]{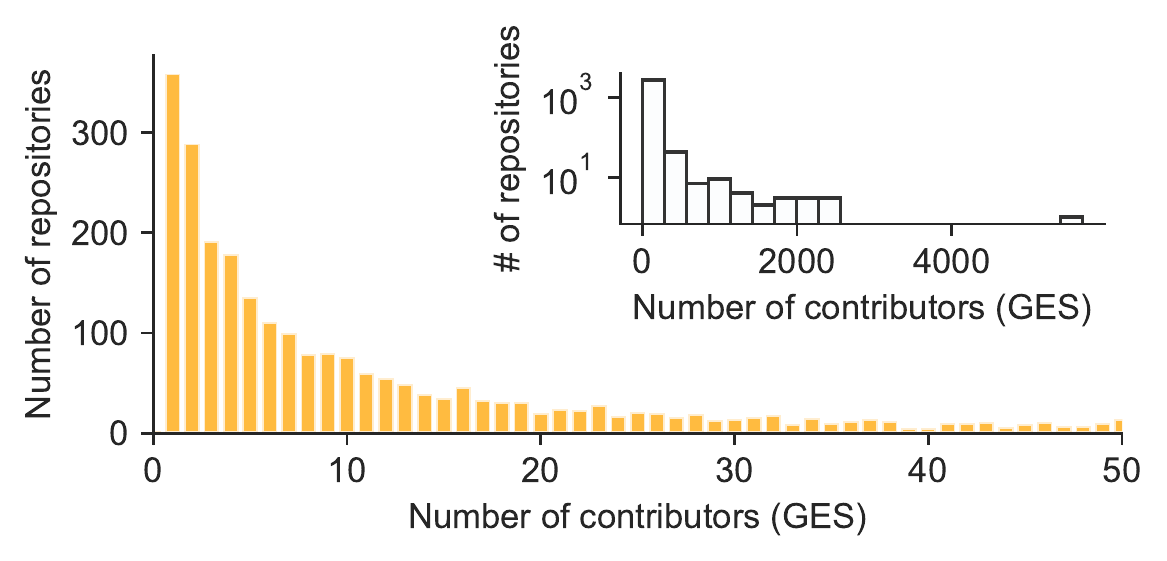}
\includegraphics[width=0.8\linewidth]{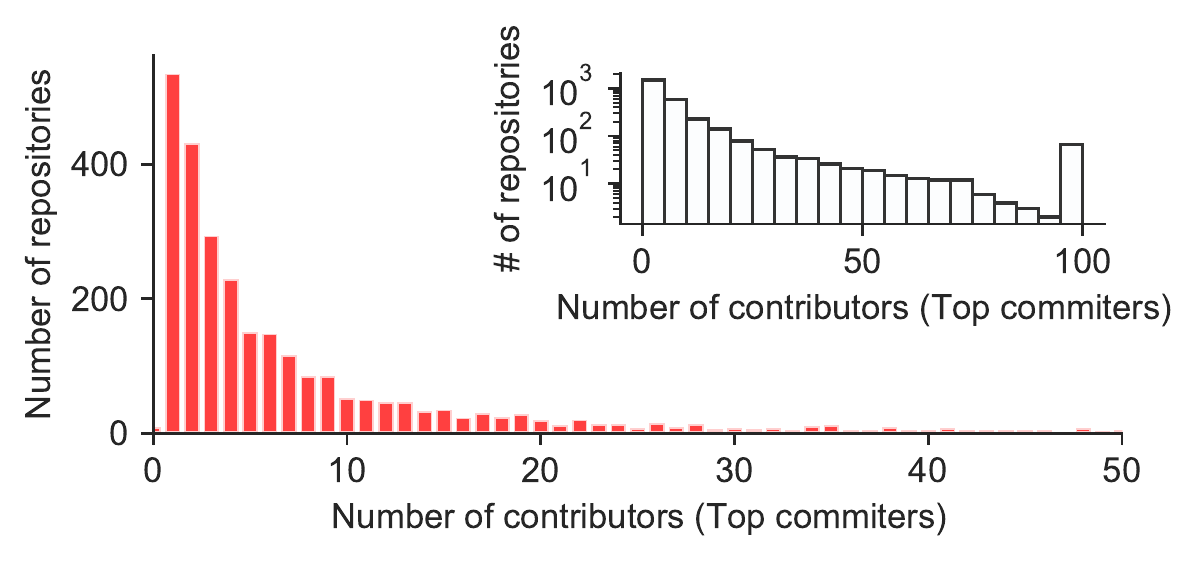}
\caption{
Number of unique contributors per project as captured by \textbf{(top to bottom)} All Contributors (AC), the Github Event Stream (GES), and the GitHub top committers.
The insets show the same histograms on a vertical logarithmic scale.
}
\label{fig:num_contrib}
\end{figure}

\subsection{RQ.1: How do models of contributorship acknowledgment differ?}
For our first research question, we want to understand the kind of work each model recognizes.
Are some models more generous? Are there contributions that are more easily recognized by all models or conversely invisible to all of them?

To start answering these questions, it is perhaps best to first visualize the distribution of the number of contributors to a project as captured by the different models (Fig.~\ref{fig:num_contrib}).
This simple exercise already reveals critical differences between models.
For instance, GitHub events (GES) capture tasks such as issues and comments, as well as all changes to the codebase, which means a greater coverage in types of contributors for each project, resulting in some projects claiming thousands of contributors.
In contrast, the Github user interface limits the number of displayed contributors to 100 and only uses commits to measure contributions, giving the appearance of far fewer contributors per project. This display distribution is saturated---indicating that several projects would acknowledge more contributors if the user interface were not capped.
Given these differences, it may be surprising that we find that the distribution of the number of contributors per project has a similar shape under the GitHub top committers and AC model, although the latter has a longer tail.

Next, examining individuals projects instead of distributions over all projects, we find that AC is used to acknowledge more contributors than what the platform highlights in the ``Contributors'' panel, in 29.5\% of the cases.
In 8.1\% of cases, the AC model also identifies more contributors than what can be derived from GitHub events.
Hence, while the absolute number of contributors recognized with AC can be quite large, on average, it is the  \emph{least generous} model of the three, in the sense that it highlights fewer contributors than GitHub events or top committers.
Some projects in our sample (6.8\%) have more top committers than contributors identified with the GES because hard forks conserve commit information that GES cannot acknowledge. 
For example, many \texttt{djangocon} projects are hard forks of \texttt{djangocon/2017.djangocon.us}, so the committers to that project appear in all the projects but not in the GES.

\begin{figure}
\includegraphics[width=0.48\linewidth]{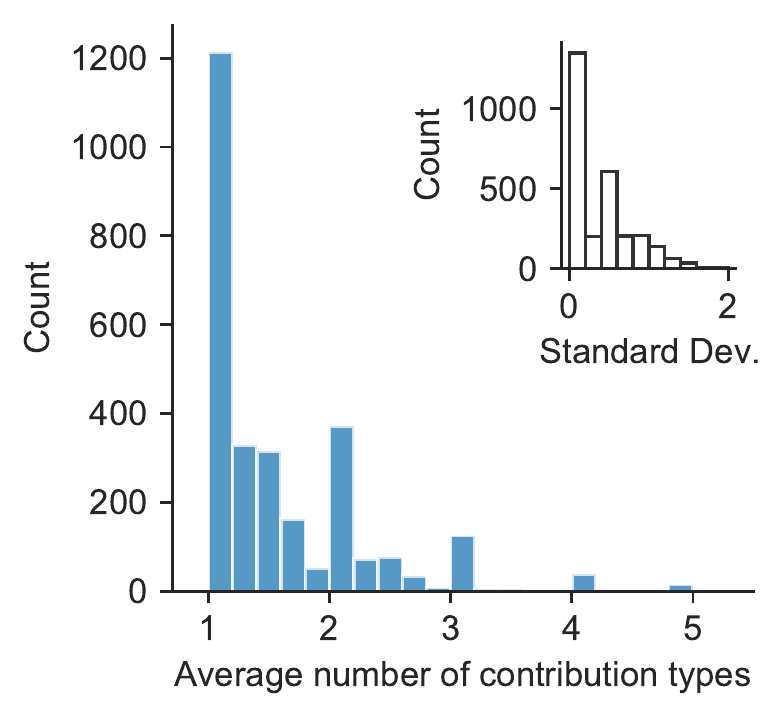}
\includegraphics[width=0.48\linewidth]{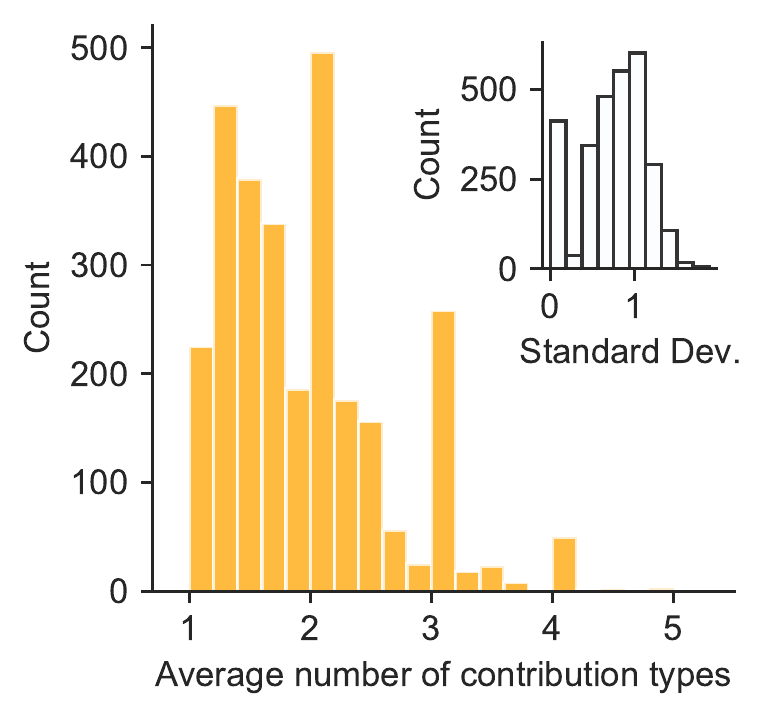}
\caption{
Distribution of the average number of contribution types per contributor, for All Contributors \textbf{(left)} and the GitHub Event Stream \textbf{(right)}.
The insets show the distributions of standard deviations of contribution types per contributor (across projects).
}
\label{fig:contrib_type}
\end{figure}

Figure~\ref{fig:contrib_type} quantifies the diversity of contribution types acknowledged in AC and the GES, the two models that identify types of contributions.
On average, most AC contributors perform about one type of task, while the GES identifies two types of contributions on average.
These quantities are tightly peaked around the mean for AC, and less so for GES.

\begin{figure*}
    \centering
    \includegraphics[width=\linewidth]{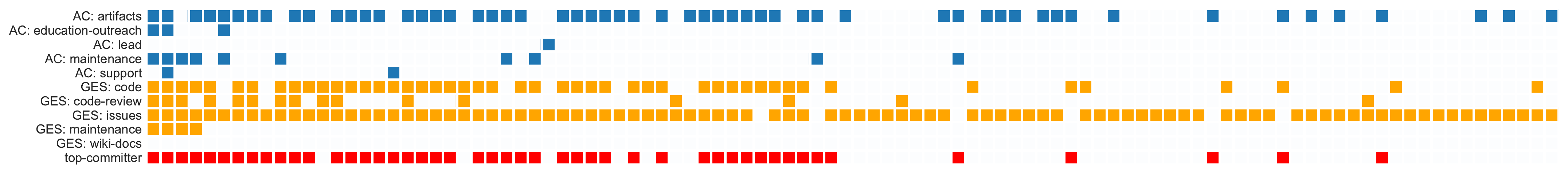}
    \caption{Example of contributorship data, here for the \texttt{all-contributors/all-contributors} repository. 
    Each column corresponds to an (anonymized) contributor and each row to a type of contribution.
    To save space, we only show the first 100 columns, ordered by the total number of acknowledged contributions across models.
    The systems used to acknowledge the contributions are: an explicit acknowledgement of contributors with the All Contributors (AC) taxonomy (in blue at the top), the automatic acknowledgement derived from parsing interactions with the repository the GitHub Event Stream (GES) (in orange in the middle), and the top committers acknowledged by the GitHub API (in red at the bottom).}
    \label{fig:example_data}
\end{figure*}

To unpack these numbers, it is helpful to analyze project-level data in greater details.
This data can be represented with an array, where one axis corresponds to people and the other to types of contribution.
We show the array associated with the All Contributors repository itself in Fig.~\ref{fig:example_data} as an example.
Upon closer inspection we find that, in this particular example, the most frequent type of AC contribution are ``artifacts''---code, plugins, doc---whereas the GES primarily reveals code contribution and code review contributions---i.e., two types of contributions.
Repeating the same exercise for all repositories and all contributors, we find that the most commonly acknowledged type of AC contribution is ``artifacts''  (22\,476 credits), followed by ``maintenance'' (5\,410 credits).
In comparison, the most commonly acknowledged type of GitHub contribution are related to issues  (113\,218 credits), followed by code contributions in the form of pull requests or push events (43\,167 credits).

We then look for significant associations between these various types of tasks.
We do this by computing the mutual information \cite{cover1999elements} of these contributions within each project.
The mutual information of contributions of types $X$ and $Y$ (say, ``artifact'' contributions acknowledged with AC and  ``code review' contributions acknowledged with the GES) can be calculated from their \emph{contingency table}.
This table records the number  $n_{00}$ of users that have done neither contributions, the number $n_{11}$ of users that have done both, and the number $n_{01}$ and $n_{10}$ that have done either of them.
We then obtain empirical frequencies $p_{ij}$ by normalizing these counts by the total number of users $\sum_{i,j} n_{ij}$ and compute the mutual information as
\begin{equation}
    \mathrm{MI}(X,Y) = \sum_{ij} p_{ij} \log \frac{p_{ij}}{u_i v_j}
\end{equation}
where $u_i = \sum_j p_{ij}$ and  $v_j = \sum_i p_{ij}$  are the marginal distributions of $p_{ij}$.
A large value of the MI$(X,Y)$ signifies that contributions of types $X$ and $Y$ are strongly associated.
Conversely, a value of 0 tells us that these types of contributions are independent.
Finally, the self-information (or mutual information of a variable with itself) equals the entropy, a measure of heterogeneity that is the smallest when all contributors contribute to that type (e.g., when all contributors code).

\begin{figure}
    \centering
    \includegraphics[width=0.9\linewidth]{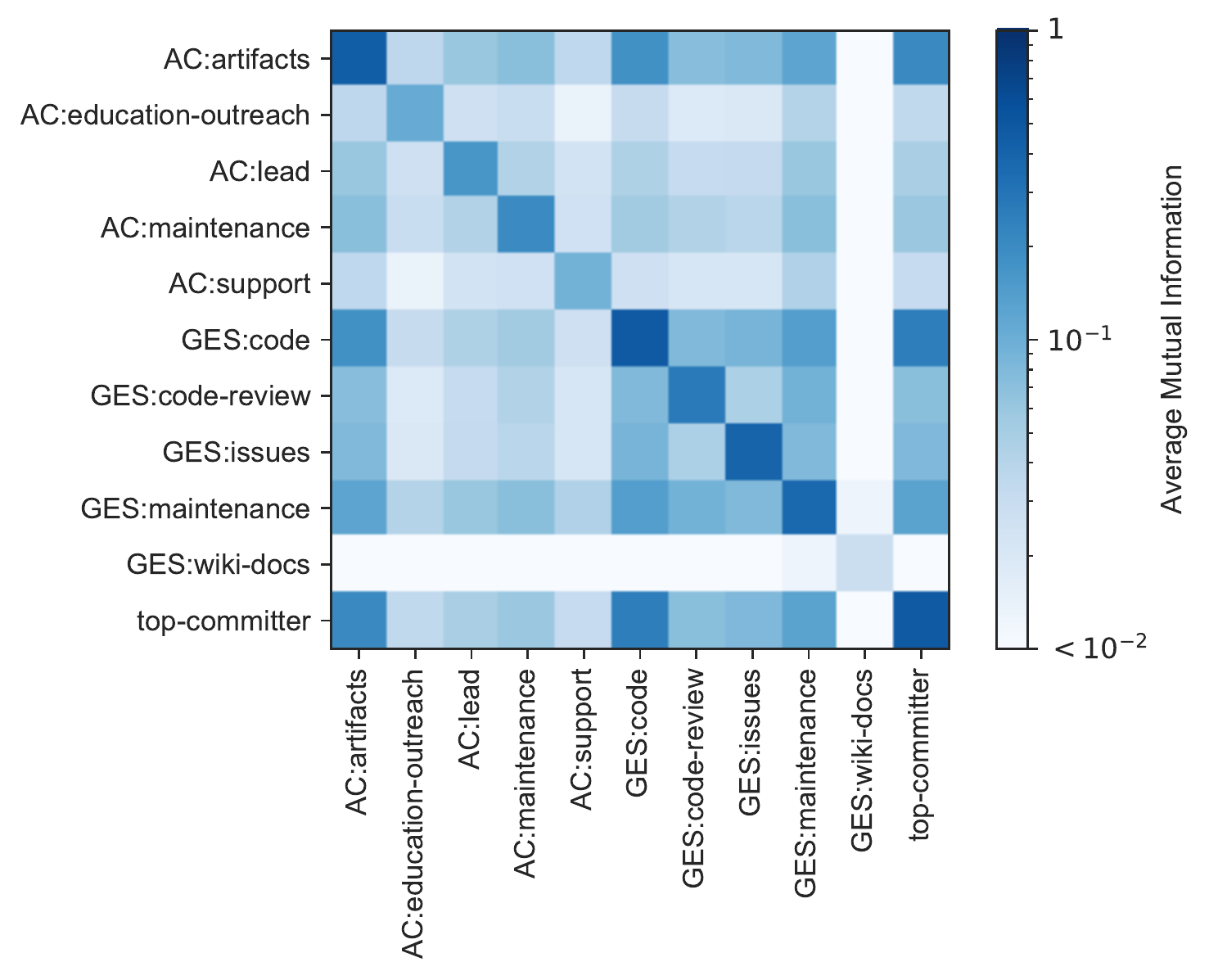}
    \caption{
        Mutual information of contribution types averaged across all repositories in our sample. 
        A large value signifies that two types of contributions $X$ and $Y$ are strongly associated.
        The diagonal shows the entropy of contribution types and is the largest when exactly half of the contributors have a contribution of that type.
    }
    \label{fig:MI}
\end{figure}

Our results appear in Fig.~\ref{fig:MI}, where we show the matrix of mutual information across contributions, averaged over all the repositories in our corpus.
The darker block-diagonal structure in the bottom right of the heatmap shows that all the GitHub contributions are strongly associated with one another, with the exception of contributions to a wiki (which are not often represented in our sample).
For example, GitHub coding contributions are associated with top committers---knowing that a user is top committer tells us that they will appear in coding events in the GES---and also artifact contributions as acknowledged by AC.
Of note, we observe that \emph{most types of AC contributions are independent of the others}, which means that models which explicitly identify categories do a good job of capturing types of contributions unrelated to code.
Strongly associated contribution types would imply that some are superfluous.

\subsection{RQ.2:  How much information is missed by focusing on repository changes as the model of contribution?}

Since different models of contributorship differ in what they acknowledge (Fig.~\ref{fig:MI}) and in terms of the total number of credited contributors (Figs.~\ref{fig:num_contrib}-\ref{fig:contrib_type}), we now ask a more specific question: How much information is missed by operating under a ``commit as contribution'' paradigm? Here, information can mean contributors beyond code (i.e., how many contributors per project have no code contributions?) but also code contributions missed because of the platform's structure (i.e., code contribution from one user committed by a separate user).

\begin{figure}
\centering
\includegraphics[width=0.45\linewidth]{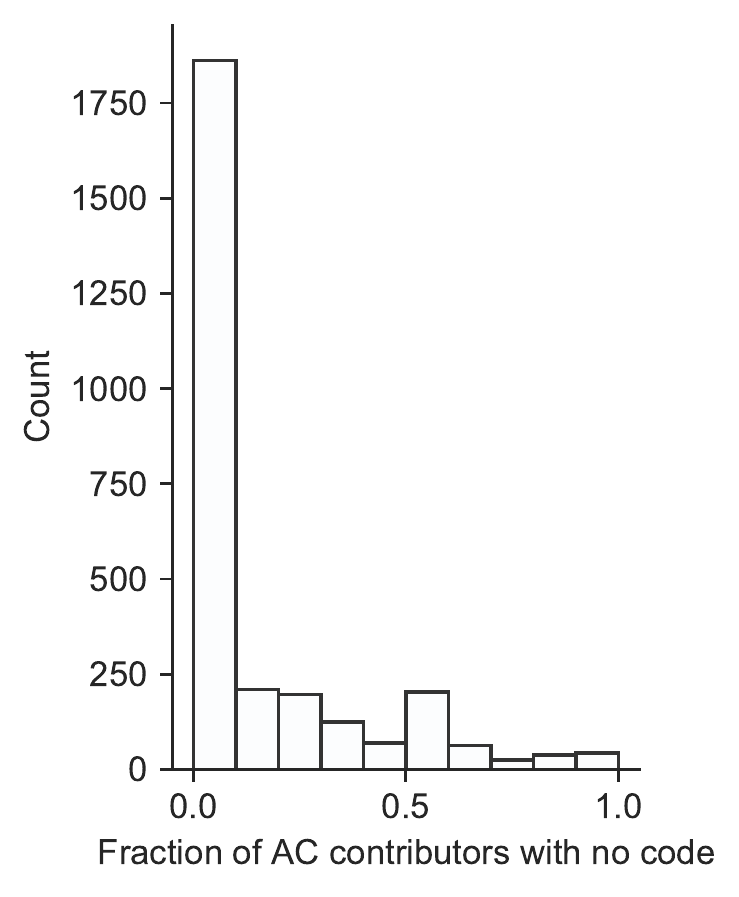}
\includegraphics[width=0.45\linewidth]{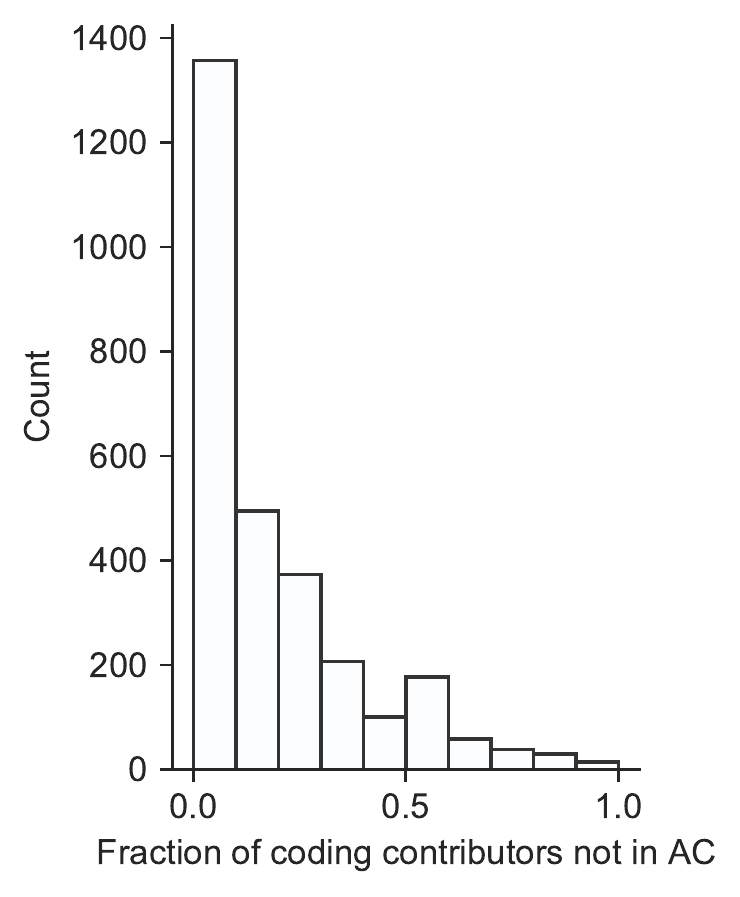}
\caption{Holes in credit attribution.
(\textbf{left}) Distribution of fraction of contributors acknowledged with All Contributors (AC) that are not top committers nor have any code contribution according to the GitHub Event Stream.
(\textbf{right})
Fraction of contributors per project that have a coding contribution (defined as being a top committer or having a code contribution as per the GitHub Event Stream) but do not appear in the All Contributors file. The histograms are complements of one another.
}
\label{fig:nocode}
\end{figure}

In Fig.~\ref{fig:nocode} (left panel), we present the histogram of projects showing the fraction of contributors acknowledged in the All Contributors file that have no code contributions according to the GES and GitHub interface.
We find that in a majority of our collected repositories, all users that appear in the AC file have at least one contribution in the form of a commit or pull request.
However, we also find that for projects with contributors beyond code (contributors that do not have a commit or pull request), there is no characteristic fraction of contributors without code contributions:
the distribution is relatively flat across repositories.

In the right panel of Fig. \ref{fig:nocode}, we show a complementary histogram, namely the distribution of the fraction of users with a coding contribution but do not appear in the AC file.
Of note, these results show that coding contributors are not a simple subset of the broader definition of contributorship considered by All Contributors.
The AC model calls for an explicit attribution of credit by a project's leadership.
Our results show that several contributors, in the strict sense that they provided code, are not acknowledged with the AC model.

\begin{figure}
\centering
    \includegraphics[width=0.9\linewidth]{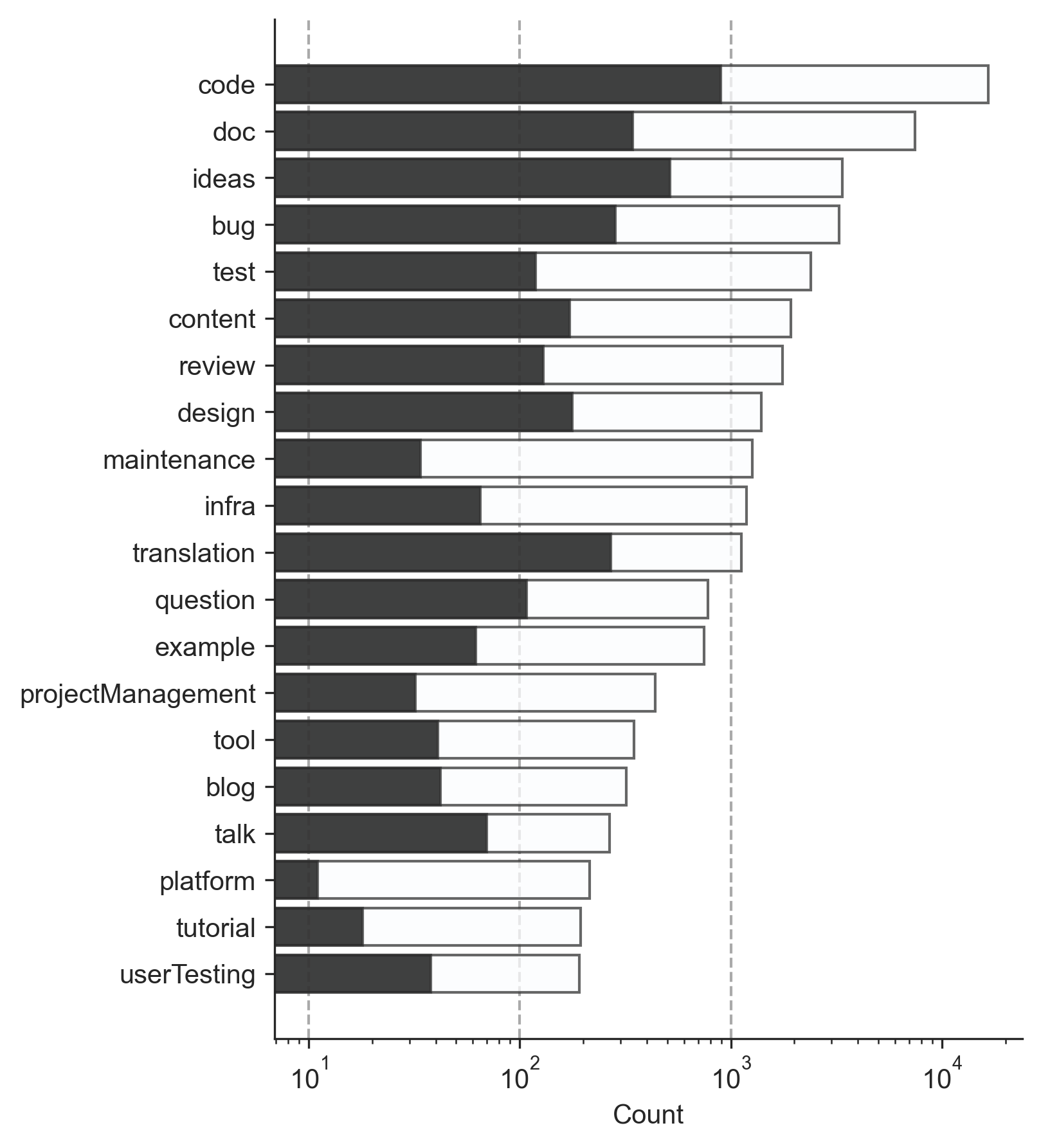}
\caption{Twenty most frequent types of AC contributions. In black, we highlight contributions made by contributors who are not acknowledged in any of the GitHub data sources (GES and top committers).
}
\label{fig:top15nocode}
\end{figure}

Figure~\ref{fig:top15nocode} investigates the nature of the AC contributions and highlights the work of contributors that are not acknowledged on the GitHub platform in any way (i.e., they do not appear in the GES and they are not top committers).
As expected, many tasks with little to do with the codebase show up in the list, such as idea generation or testing.
Somewhat surprisingly, we find at the top of the list code-related contributions such as code, documentation, or bug-finding.
These results can be explained by a mixture of mechanisms: some organizations maintain AC files that are not in sync with where the work occurs; the workflow of some OSS projects involves mirroring code on GitHub, which masks code-contribution events;
and our filters could miss duplicates created by pushing a clone or near-copies of a repository without altering the AC file.  
In all these cases, developers might be acknowledged for coding work not captured by the platform.

For our current research question, these results stress that, by focusing on direct code contributions like commits and pull requests, most of the contributions missed could be indirect code contributions such as development occurring outside of our sample (i.e., outside of GitHub).

\begin{figure}
\centering
\includegraphics[width=\linewidth]{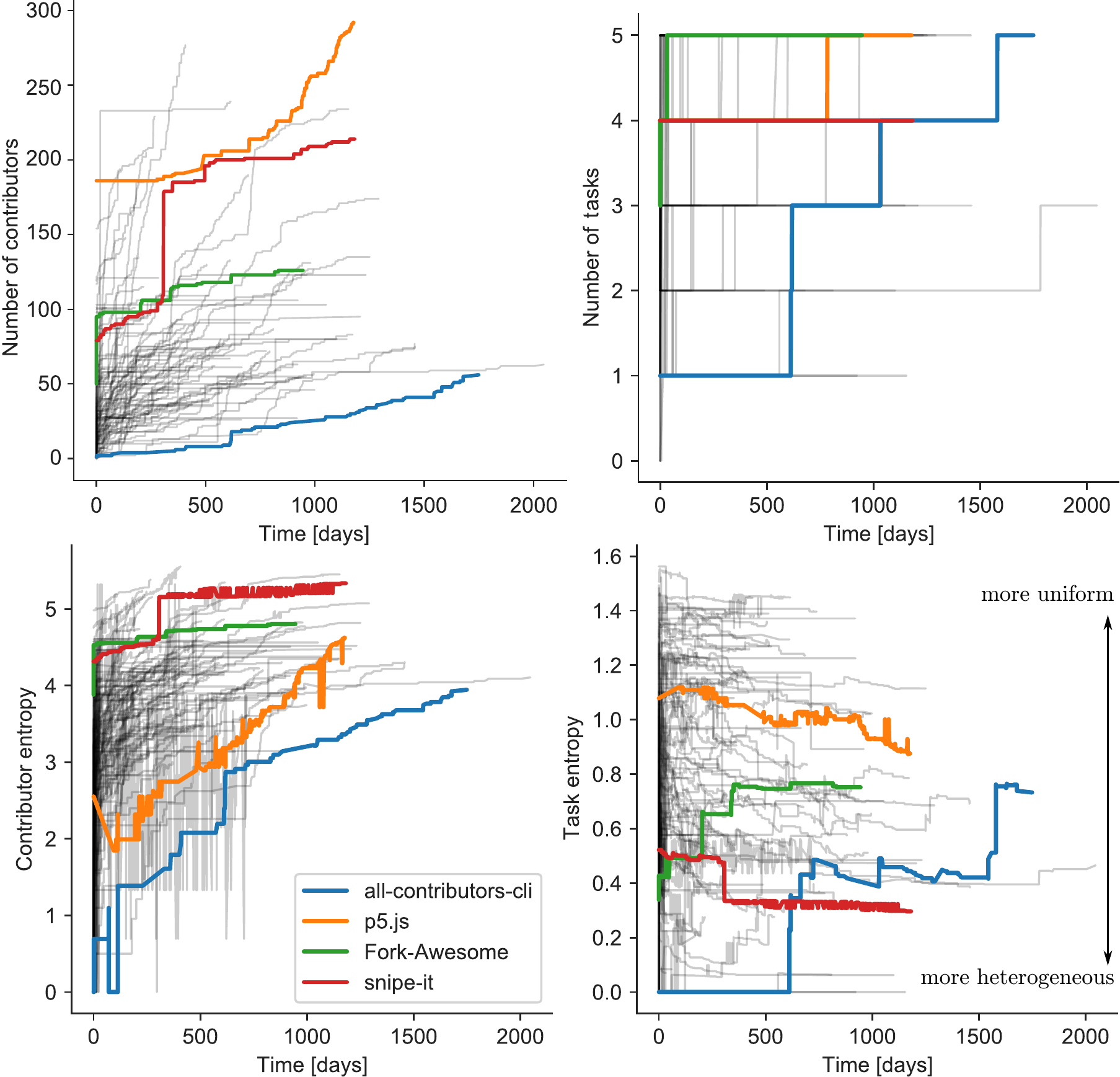}
\caption{Evolution in the All Contributors file of the largest 100 projects of \textbf{(top left)} the number of contributors, \textbf{(top right)} the number of tasks, \textbf{(bottom left)} the contributor entropy and \textbf{(bottom right)} the task entropy. Recall that a low contributor entropy means that a few users carry out most of tasks types and that low task entropy means that a few tasks dominate the contributions.
Conversely, high entropy implies a more uniform distribution of contributions over users or tasks.}
\label{fig:growth}
\end{figure}

\subsection{RQ 3: How do contribution to open source projects evolve as they age?}

Shifting our focus away from how acknowledgment models differ, we can also use the contributorship data to understand more about how OSS is made.
To this end, we first investigate the role of time and how it affects the makeup of contributors to a project.
We first look at the growth of projects over time in the top row of Fig.~\ref{fig:growth} (here defined solely through their All Contributors file), where time is defined as the number of days since the creation of the AC file.
find periods where projects are relatively static or undergoing linear growth and periods of faster (e.g., exponential) growth.
Hence, a given project's state of contributors is not in equilibrium but should be studied as a dynamical process.

\begin{figure*}
\centering
\includegraphics[width=0.75\linewidth]{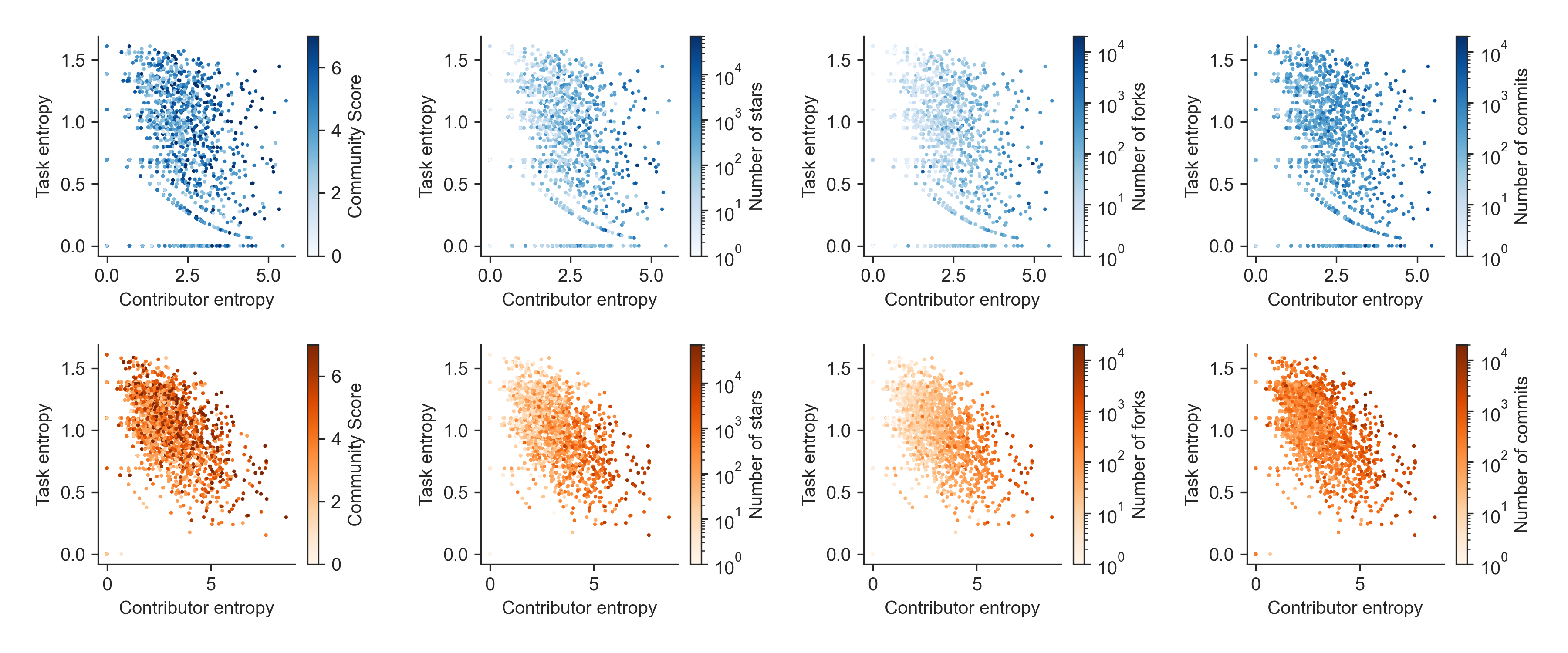}\quad
\includegraphics[width=0.2\linewidth]{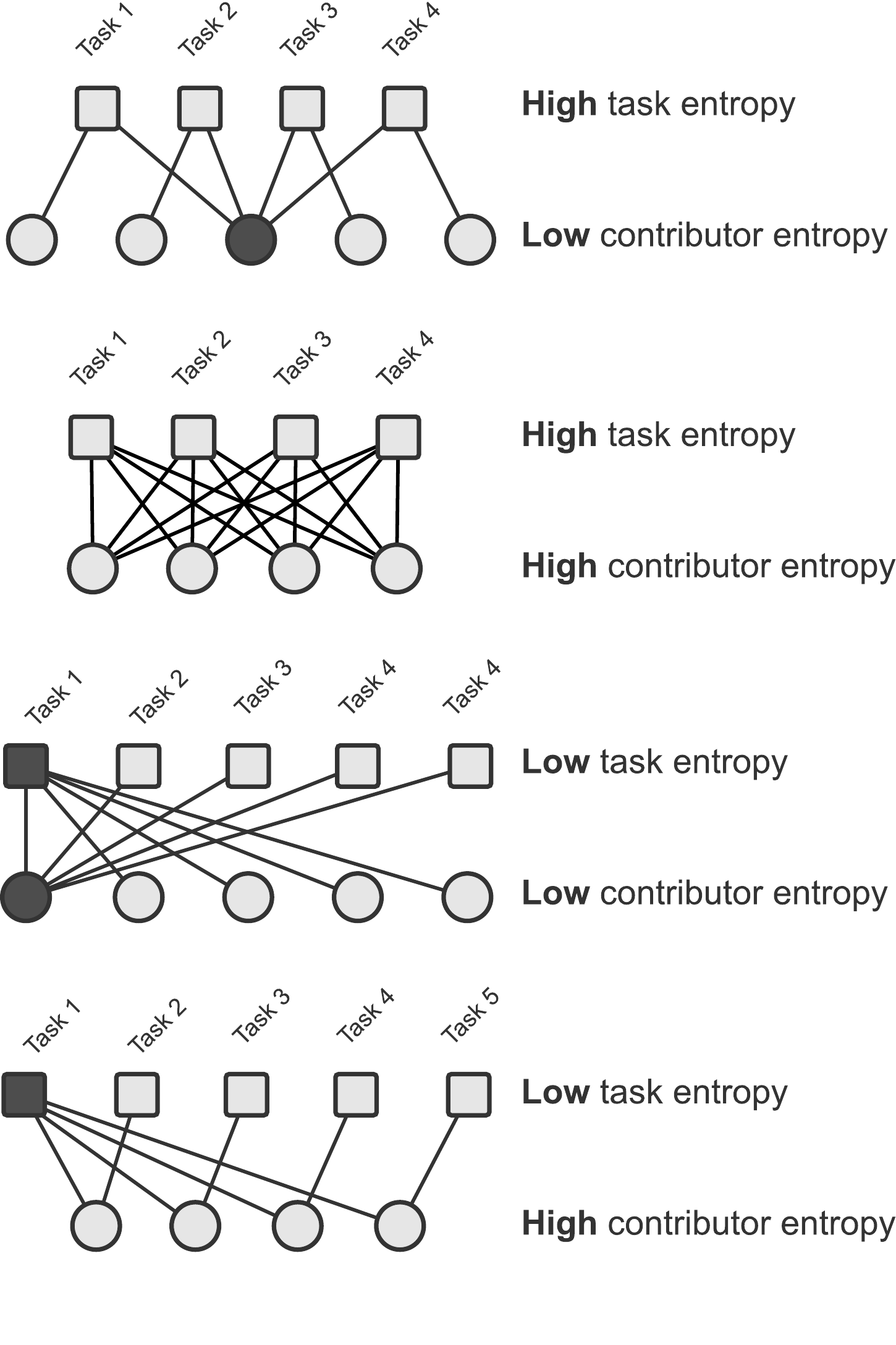}
\caption{
We measure the heterogeneity of contributions to different projects (unique markers in each plot) by calculating the entropy of tasks (contribution types) per contributor and contributor per task (Eq.~\ref{eqn:entropies}).
A high contributor entropy suggests a uniform distribution of task types over contributors, and a high task entropy suggests a uniform distribution of task types over contributions.
Conversely, high entropies suggest a concentration of efforts upon a few contributors/tasks.
\textbf{(top row)} Map of all projects in this entropy space using contributions characterized by All Contributors; \textbf{(bottom row)} equivalent map using the GitHub Event Stream.
The four columns look at correlation in this space with different repositories: \textbf{(left to right columns)} GitHub's community score, stars, forks, and commits.
Examples of projects with different combinations of high/low task/contributor entropies are shown to the side, with squares representing contribution types and circles representing contributors.
\label{fig:space}
}
\end{figure*}

To understand how the distribution of contributions evolves over time, we then look at a project's trajectory with what we call the task entropy and contributor entropy of the project.
We define these entropies as:
\begin{equation}
\label{eqn:entropies}
  H_{\mathrm{contrib.}} =  -\sum_{i} \frac{k_i}{m} \log \frac{k_i}{m}, \quad  H_{\mathrm{task}} = -\sum_k \frac{q_j}{m} \log \frac{q_j}{m},
\end{equation}
where $k_i$ is the number of contribution types (or, tasks) made by person $i$, where $q_j$ is the number of people doing contributions of type $j$, and where $m=\sum_i k_i=\sum_j q_j$ is a normalization ensuring that $\{k_i/m\}_i$ and $\{q_j/m\}_j$ are probability distributions.
The contributor entropy becomes larger the more uniformly distributed tasks are between all contributors.
Likewise, the task entropy is the largest when contributors are distributed uniformly between all tasks (i.e., contribution types), see Fig.~\ref{fig:space} (right) for a schematic illustration of extreme cases.
The entropy~\cite{cover1999elements} has been used before to quantify the diversity of contributors to files~\cite{casebolt2009author,vasilescu2014variation}, effectively asking which whether a file is the work of many developers or only a few; here we use it to ask questions about tasks and contributors.

In the bottom row of Fig. \ref{fig:growth}, we show the separate evolution of task and contributor entropies over time for the 100 projects with the most AC contributors. 
The time series of contributors are consistent with what we expect from projects whose number of contributors all grow over time: Contributor entropy also grows over time, resulting in an increasingly homogeneous distribution of tasks over contributors. 
This change could be due to increased project complexity or simply a higher drive to recognize the teams' diverse work as projects mature.

Surprisingly, we find no such general trend in the time series of task entropies.
For many projects, the distribution of contributors over tasks seems to stabilize at drastically different values for different projects or slowly become more heterogeneous over time. Some time-series also feature sudden drastic jumps, see \texttt{all-contributors-cli}, likely related to internal decisions to credit new types of contributions as shown in Fig. 9.

\subsection{RQ.4: Can we classify projects based on patterns of contributions?}

As a final research question, we ask whether patterns of contributions can help us find archetypal projects.
To this end, we use the entropies introduced for RQ3 and define a space of projects, where each project is represented by a contributor entropy and task entropy that is computed with the most up-to-date contributorship information available.
For this question, we again use the entire sample of 2855 projects.
We show the resulting space in Fig.~\ref{fig:space}; the columns highlight the relation between the entropies and various covariates, namely the community score, the number of stars, forks, and commits.
The top row shows the space obtained when we use AC contributions to compute the entropies, and the bottom row shows the same space but using GitHub events.
Both models of contribution acknowledgment span a similar range of task entropies, but the GES leads to a broader range of contributor entropies.

One can think of the corners of these spaces as corresponding to different archetypal projects.
High contributor entropy and high task entropy indicate that every contributor does every type of contribution---an unlikely situation unless the project is very small.
Low contributor entropy and low task entropy mean that few or one contributor does all the types of contributions and that there are not many types of contributions being made.
Again this tends not to happen unless a project comprises only code and is developed by a very small team.
Mixed situations are more typical, and, indeed, projects tend to organize along an anti-correlated diagonal axis in Fig.~\ref{fig:space}.
Low contributor entropy and high task entropy (top-left corner) occur when a few contributors do all types of tasks (a pattern of contributions we have seen in young projects, see Fig.~\ref{fig:growth}).
Conversely, high contributor entropy and low task entropy (bottom-right corner) correspond to a situation where all the contributors have similar contribution patterns, mainly concentrated around a few tasks.
It seems to be the purview of more established projects, where most contributors create code (which is, after all, the primary product of most OSS projects).

Turning to the connection between entropies and various covariates, we find that a high contributor entropy is strongly associated with higher popularity (measured with stars and forks) and modestly associated with a more extended project history (measured in commits).
For instance, the  Pearson's correlation coefficient of the GES contributor entropy with the logarithm of the number of forks (plus one) is $r=0.87$, while the $r=0.56$ for the logarithm of the length of the project history. 
The results are consistent with Fig.~\ref{fig:growth}, as old and, on average, more popular projects tend to have higher contributor entropy. 
However, we also find no strong relation between entropies and community scores ($r\in[0.1,0.23]$ for the correlation of the entropies values with the community scores).
This lack of correlation is somewhat surprising, as we might have expected compliance with the community profile to be a strong indicator of other prosocial behavior, like having a well-balanced (i.e., high entropy) contributor roster.

\section{Discussion}
\label{sec:discussion}

We found in many ways that contributions listed by a project's All Contributors (AC) file were less ``generous'' than contributions extracted from the project's GitHub Event Stream (GES), with fewer types of contributions per user and over 90\% of projects containing fewer acknowledged contributors in their AC file than those appearing in the GES.
At the same time, we have found, perhaps unsurprisingly, that AC acknowledges work that could not have been recognized at all otherwise.
So, why is AC less generous even though it can acknowledge more contributions in theory?
One reason may be a higher threshold for inclusion: a project may be (implicitly or explicitly) using their AC file to denote what they consider to be \textit{substantive} contributions.
Another reason might be that the manual labor involved in acknowledging and vetting contributors acts as a barrier to entry, limiting the number of recognized contributors.

While OSS roles are not limited to software development, and AC's contributor taxonomy specifically accounts for non-code contributions, we still observe that code contribution is the dominant contribution type: most users listed in a project's AC make a code contribution. 
Given that OSS teams tend to be small \cite{krishnamurthy2002cave} and only larger teams tend to utilize and benefit from non-code support~\cite{eghbal2020working}, maintaining AC files may be helpful for only those larger teams.
Similarly, most teams feature a small core of developers who perform most code work~\cite{ducheneaut2005socialization}, and AC acknowledgments may be less likely to reach contributors beyond that core.

All of these observations suggest that the AC model operates on a stricter yet fuzzier model of contributorship; what is and what is not a ``significant'' contribution is largely up to a project's team, and the barrier to entry seems large.

The differences between models of contributorship acknowledgment have implications for diversity, equity, and inclusion (DEI) in OSS~\cite{opensourcesurvey,izquierdo2018openstack}.
For example, women enter the OSS community later and for different reasons than men and tend to participate less if they have children.
Women are also more likely than men to perform non-coding tasks,
and so efforts such as AC are crucial to more equitably acknowledging efforts and helping projects be more inclusive.
At the same time, the stricter definition of contributorship implemented in models like AC might also be at odds with DEI efforts.
Studying, for example, gender effects within these data is thus an essential avenue for future research.

Our findings also have implications for the validity of analyzes of software repositories.
First, our results show that when one mines software repositories to conduct a sociological analysis of teams, data sources like the GES will likely include contributors that do not meet the implicit standards team members hold each other to.
At the same time, the GES may miss key contributors who do not work on code because GES contributions only correlate with artifacts (Fig.~\ref{fig:MI}).
The GES is thus likely to simultaneously over- and under-estimate the breadth of contributions made to a project.
Second, our results also show taxonomies should not be used if the goal is to determine the projects' footprint, defined as the total number of developers having interacted with projects (as GES does). 

And while we have found disparities between models, we also found surprising similarities.
Of note, we found that the two models of credit attributions paint a similar picture of a project's lifespan: contributors distribute homogeneously across contributions types as project mature---whether we measure maturity in terms of popularity (Fig.~\ref{fig:space}) or time (Fig.~\ref{fig:growth}).

The All Contributors project paves the way towards a standardized acknowledgment of contribution in OSS~\cite{alliez2019attributing}.
With broader identification and attribution of the work critical to releasing and maintaining software, there is hope for both more effective and more inclusive OSS communities.
By standardizing contributorship, researchers can begin to measure aspects of OSS that go beyond the code itself.
Yet a standard taxonomy is not enough, even if adopted broadly: it must also be adopted consistently.
If communities define and apply criteria for using the taxonomy differently, then cross-comparison becomes challenging (see also Sec.~\ref{sec:threats}).
This was already an issue for the modest corpus we have analyzed, as we have encountered several mutations to the AC model while gathering our dataset. 


\section{Threats to validity}
\label{sec:threats}

\begin{figure}
    \centering
    \includegraphics[width=0.8\linewidth]{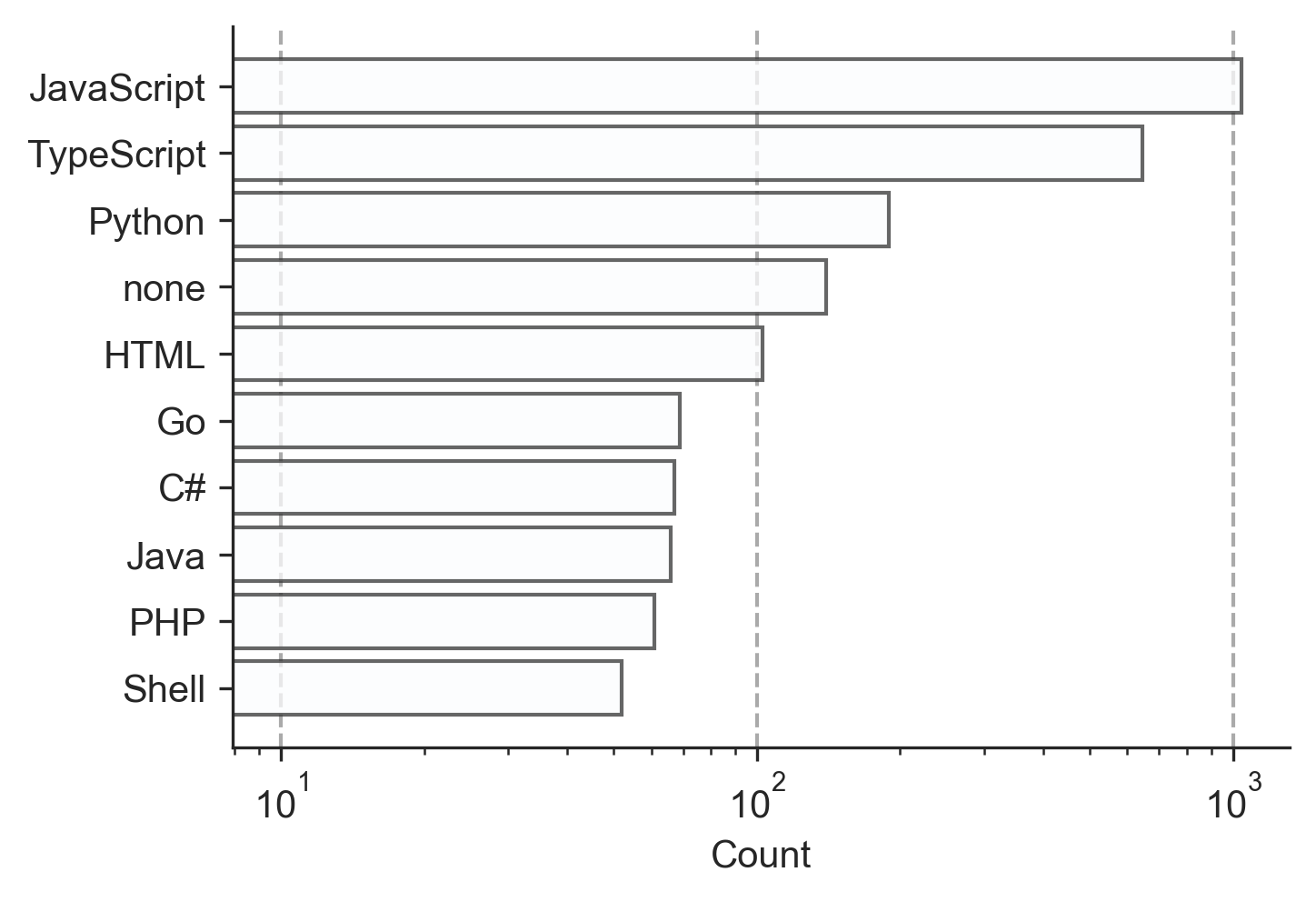}
    \caption{Primary programming language of the repositories in our sample (top 10) on a logarithmic scale.
    We note that a large fraction of the repositories implementing the AC model are written in  JavaScript (36.6\%) or TypeScript (22.7\%), a superset of JavaScript).
    A few Python  projects (189) have also implemented the AC model.}
    \label{fig:languages}
\end{figure}

Perhaps the largest threat to the validity of our results stem from the fact that we have used a sample of convenience---the few thousands of projects that implement the All Contributors model of contributorship acknowledgement on GitHub.
This has several implications.
First, our sample is biased towards projects willing to go out of their way to acknowledge contributors using a method that is not widely used, which is a particularly prosocial behaviour.
Hence, the group of acknowledged the contributors might be particularly broad relative to a typical OSS project.
This effect might be compounded by the fact that the All Contributors model is most popular in the JavaScript community (see Fig.~\ref{fig:languages}), which is known for its particularly collaborative practices, and for its focus on developers rather than projects  \cite{eghbal2020working}.
Second, the code that appears on GitHub is itself a skewed sample of OSS software so our results might not generalize to other services or populations \cite{furtado2020successful}.

We have also assumed that logins correspond to a single user.
This is a relatively safe assumption since we use platform-level data (i.e., GitHub logins) to identify users rather than git logs~\cite{fry2020dataset,amreen2020alfaa} or mailing lists~\cite{wiese2016mailing} where aliases are poorly resolved.
Beyond aliases we have assumed that login information is consistent across data sources (AC and GES).
This assumption is unlikely to lead to major issues, since \texttt{.all-contributorsrc} files are typically managed by a bot via comments on pull requests and issues, at the platform level, where users are again identified by their GitHub login.
Nonetheless, in some instances All Contributors files are handled manually, and we have encountered 7 instances of names provided without logins (for which we assumed there was no matching code contribution or interactions with the repository).

On a technical note, we have filtered bots using simple pattern matched on names combined with a hand-crafted list of bots.
Nonetheless, the filter might be non-exhaustive, in which case some our results could conflate the activity of bots with that of human developers.
More exhaustive filtration techniques could reduce this threat in the future~\cite{dey2020detecting}.

Another threat to validity is the fact that repositories are not projects  \cite{kalliamvakou2014promises}.
We have gone to great length to mitigate this threat by filtering repositories heavily as to keep only the ones where development is or was active.
However, there are still cases where our filters cannot do much, especially when organizations control several repositories.
It may be the case that organizations concentrate all their contributors within a single All Contributors file but distribute work across several repositories (or vice-versa).
Further, non-code contributions may be hard to assign---should ``finance'' appear in all the repositories of an organization or only the main one?
The lack of clear signaling around these norms mean that usage may vary in practice
and implies that OSS needs further standardization for how contributions are acknowledged.

Finally, while our sampling strategy and mining goals allow us to avoid most of the perils usually involved with mining GitHub (like the lack of activity for most repositories, that few projects use pull requests, or that some repositories are private or non-software projects)~\cite{kalliamvakou2016depth}, we may be affected by some issues specific to git~\cite{bird2009promises}.
For instance, the history of git projects is revisionist, which implies that the GES  might not give us a fully accurate picture of development \cite{bird2009promises}.

\section{Additional related work}

Early studies of contributions in software development had little to do with credit categorization and attribution, instead primarily investigating the nature of software development itself.
For example, Glass \textit{et al.}~\cite{glass1992software} aimed to quantify to what extent contributions involve  ``human ingenuity'' (or ``intellectual tasks'') as opposed to ``routine procedures'' (or ``clerical tasks'').
To do so, they rely on small-scale experiments (e.g.
six graduate students in a software development course) and on a complicated taxonomy developed by Henderson \& Cooprider~\cite{henderson1990dimensions} composed of 61 so-called intellectual tasks and 22 clerical tasks.

More recent studies tend to leverage large-scale data available on OSS.
These efforts are often focused on committers, whose contributions are directly available in repositories or on specific communities with available lists of members (e.g., the Apache Software Foundation \cite{squire2013project}).
Other efforts focus on different types of contributions, such as Hamasaki \emph{et al.}~\cite{hamasaki2013does} who present case studies and datasets on code review.
While most of their work evaluates the quality of reviews and individual performance, they also analyze the evolution of different roles in OSS development as categorized by the Android Open Source Project documentation\footnote{\url{https://source.android.com/setup/start/roles}}.

These large-scale datasets have been used to measure or categorize significant contributions to software development. To this end, there exist a series of models developed to use software repository (quantified through lines of codes metrics) as well as mailing list or wikis (quantified through threads or pages) to estimate effort \cite{amor2006effort} or measure contribution \cite{gousios2008measuring}. Interestingly, unlike in our current paper, Gousios \emph{et al.}~\cite{gousios2008measuring} also includes the possibility of negative contributions such as a line of code that introduces a bug, commits without comments, or reporting invalid bugs.

With these large datasets and new analysis tools comes the need to better acknowledge and quantify contributions. 
In that spirit, Capiluppi~\emph{et al.}~\cite{capiluppi2012developing} proposes to quantitatively credit developers through h-index like metrics. 
These proposals do not consider the different types of contributions or relative differences in the scale of contributions.
Therefore, studies like Lima \emph{et al.}~\cite{lima2015assessing} have exposed the need for more refined models of contributions that embrace contributions of different natures as well as task distributions within projects.

Notably, Alliez \textit{et al.}  \cite{alliez2019attributing} recently reported on their experience on crediting software within Inria, the French institute for digital sciences.
They recommend rich taxonomies for software contributions with qualitative scales to enhance the visibility and impact of research teams.
Ramin et al. \cite{ramin2020more} also recognized the need for contributorship taxonomies that go beyond code and propose to use scrum roles.
Milewicz and collaborators also carried a survey-based analysis contributorship in scientific software (on 72 developers) to empirically informed better models of the development process \cite{milewicz2019characterizing}.

In the academic context, taxonomies such as the CRediT framework~\cite{holcombe2019contributorship,holcombe2019farewell,brand2015beyond,allen2019how} have allowed important analyses of labor divisions \cite{lariviere2020investigating}. Similar work could be enabled in software development through new contributorship taxonomies. 

\section{Conclusion}

That software development is driven by elements beyond code is well accepted, with the social, financial, and educational dimensions of OSS having attracted particular attention. Models of contributorship acknowledgment embrace this central aspect of OSS, and aim to give an objective and holistic acknowledgment of contribution across all these dimensions.

Our results characterize the main model of contributorship acknowledgment by examining thousands of projects adopting the All Contributors model and comparing these with default platform statistics.
We find that while community-generated attributions make non-code related work more visible, it also tends to be less generous and can miss several contributors to a project.
We hypothesized that it could be due to contributions deemed less significant by the community, but which still should not be invisible if we are to capture a complete picture of OSS projects.
Future qualitative work will help shed light on such trade-offs and on why projects are compelled to implement models like All Contributors.

Our findings stress the importance of defining new contribution taxonomies, or refining existing ones, through workshops to accommodate diverse projects in OSS.
This collective development is critical for both large-scale adoptions of taxonomy and its uniform implementation.
Clearer, more complete taxonomies can help us better observe OSS projects, both by highlighting meaningful contributions and by bringing often invisible work to light.

\section*{Acknowledgments}

This work is supported by Google Open Source under the Open Source Complex Ecosystems And Networks (OCEAN) project.  
Any opinions, findings, and conclusions or recommendations expressed in this material are those of the authors and do not necessarily reflect the views of Google Open Source.


\end{document}